\documentclass[pmlr]{jmlr}
\usepackage{booktabs}
\usepackage{longtable}
\usepackage[load-configurations=version-1]{siunitx}
\usepackage{multirow}

\jmlrvolume{}
\jmlryear{2026}
\jmlrworkshop{Impactful and Responsible AI Systems for Education}

\title[EvalConvoLearn]{EvalConvoLearn: An Open-Source Framework for Evaluating Grounded Learner Simulations in Tutoring Conversations}

\author{\Name{Baptiste Moreau-Pernet}
\Email{baptiste@levi.digitalharbor.org}}

\begin{document}

\maketitle

\begin{abstract}
Conversational learner simulations are valuable tools for testing learning theories, evaluating instructional materials and automated tutors, or powering teachable agents. Recently, large language models (LLM) have enabled richer, more naturalistic interactions with simulated learners; however, no open framework exists for evaluating whether such simulations faithfully reproduce real learner behavior. We introduce \textbf{EvalConvoLearn}, an open-source framework that assesses learner simulations along two axes: \emph{learning behavior} (skill-conditioned mastery outcomes) and \emph{conversational quality} (talk moves, error type distributions, question rate, turn length). EvalConvoLearn measures how closely a simulated learner approximates answer distributions observed in data by grounding metrics in authentic tutoring conversation datasets, and anchoring generated tutor responses in existing tutor utterances. The framework is demonstrated on a dataset of tutoring dialogues, including results for two LLM-based learner simulations, and the published GitHub\footnote{\url{https://github.com/RenaissancePhilanthropy/EvalConvoLearn}} code.
\end{abstract}


\section{Introduction}
\label{sec:intro}

Simulated learners serve diverse purposes in education \citep{koedinger2015}: testing theories of learning and instructional approaches, automating authoring of tutoring tools, or acting as teachable agents through which students learn by teaching. Early work used cognitive models---e.g., Betty's Brain~\citep{leelawong2008designing} and SimStudent~\citep{matsuda2007predicting}---to simulate learner interactions in structured tutoring environments. More recently, LLM-based tutors interact with students through naturalistic conversations (e.g., Khan Academy's Khanmigo,\footnote{\url{https://www.khanmigo.ai/}} Google LearnLM\footnote{\url{https://blog.google/outreach-initiatives/education/google-learnlm-gemini-generative-ai/}}). Evaluating such tutors at scale requires correspondingly capable \emph{conversational} learner simulations.

However, ensuring that such simulations are valid and useful for the various goals above is an open problem: simulating realistic student learning and conversational characteristics is complex because LLMs are trained as helpful assistants \citep{martynova-etal-2025-llms}. Previous work has assessed simulators on their ability to match real-world learner skill mastery over time \citep{maclellan2016apprentice} or reproduce learner conversational characteristics \citep{perczel2025teachlmposttrainingllmseducation}. \citet{Scarlatos2025} introduce a consolidated set of metrics by matching a simulated learner response with the real student message given the conversation history.  However, they only evaluate simulations on one turn and not at the dialogue level, and their work is closed-source, making reproduction and extensions to other datasets difficult.

We present \textbf{EvalConvoLearn}, an open-source framework evaluating the quality of conversational learner simulations. It can serve as a foundation to benchmark learner simulations across educational contexts and design goals for learners, including error types, learning trajectories, or social and interaction metrics \citep{yuan2026validstudentsimulationlarge}. EvalConvoLearn scores simulations' ability to produce realistic student behavior \emph{distributions} across metrics extracted from tutoring conversation datasets.

\section{The EvalConvoLearn Framework}
\label{sec:framework}

\begin{figure}
    \centering
    \includegraphics[width=0.8\columnwidth]{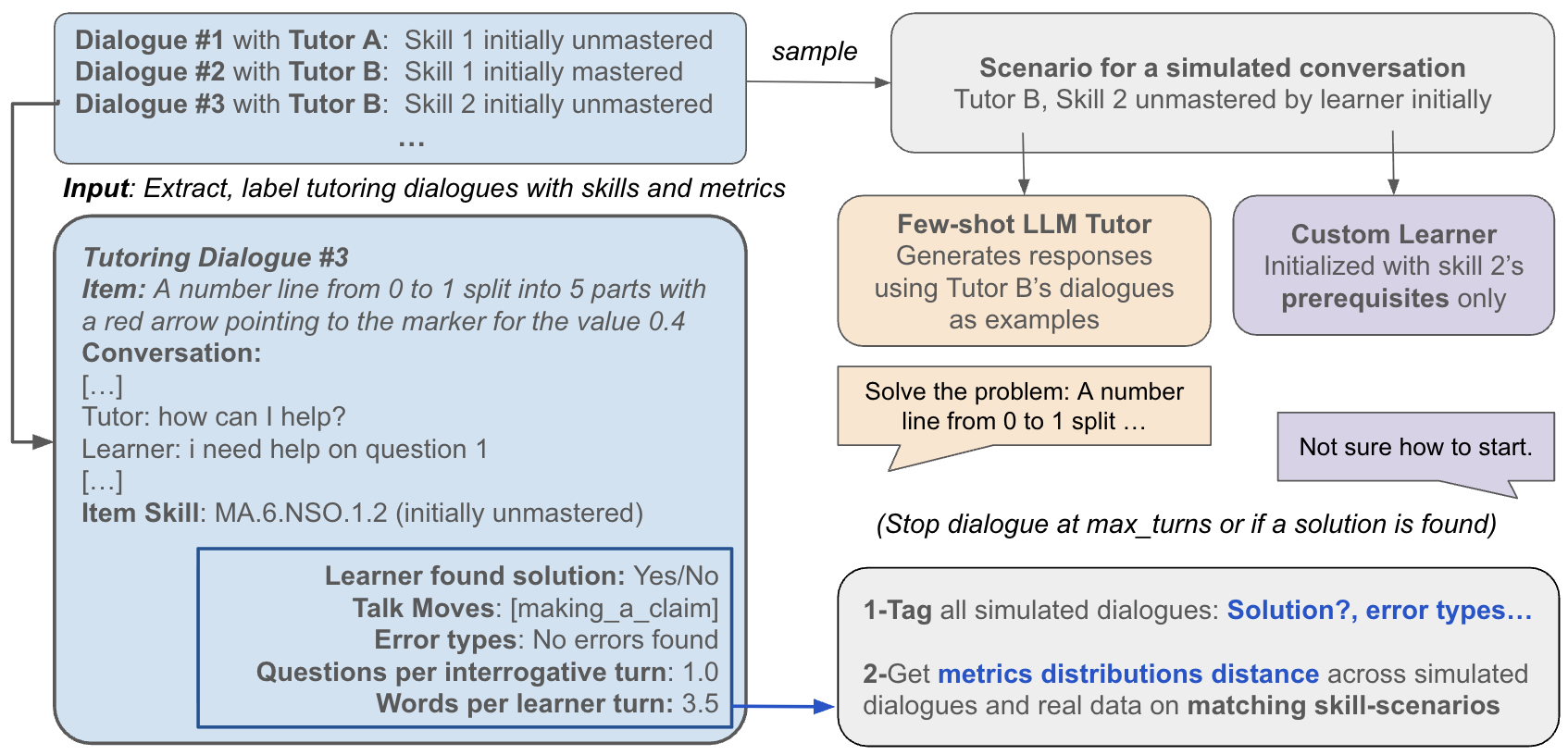}
    \caption{EvalConvoLearn learner evaluation steps.}
    \label{fig:evalconvolearn_steps}
\end{figure}

EvalConvoLearn evaluates a learner simulation along two axes: (i) appropriate skill acquisition \citep{weitekamp2025tutorgym} and (ii) conversational realism. The metrics are intentionally simple and target learner goal-aligned fidelity rather than surface realism, following \citet{yuan2026validstudentsimulationlarge}.
Given a tutoring conversation dataset with practice items and tagged skills (or items automatically tagged), EvalConvoLearn extracts \emph{learning scenarios} defined as the product of a target skill and the learner's prior mastery of it (Figure~\ref{fig:evalconvolearn_steps}). For each scenario, the framework runs simulated problem-solving conversations between the custom learner and EvalConvoLearn's tutor, then compares simulated and real metric distributions. The tutor is grounded at the individual tutor level with few-shot dialogue examples from other conversations in the dataset to simulate more realistic responses.

\subsection{Learner Simulation Interface}
\label{sec:interface}

EvalConvoLearn defines a minimal interface of four functions that any custom learner simulation must implement to be evaluated: The learner \textbf{initializes} its knowledge state according to a skill profile (mastered/unmastered skills or prerequisites). Then, it \textbf{generates a response} given a conversation history and its knowledge state, and is able to \textbf{update} its knowledge state from a conversation. Finally, it should be able to return whether it currently \textbf{masters} a given skill (used to match learners with mastery scenarios).
If a custom learner cannot look up specific skills in its knowledge state, a default initialization loop runs problem-solving conversations along the skill's prerequisites tree, while assessing progress at each step via skill-aligned items using the learner's \textit{response generation} functionality.

\subsection{Scenario-Based Metrics Evaluation}
\label{sec:scenario_eval}

\begin{figure}
    \centering
    \includegraphics[width=0.9\columnwidth]{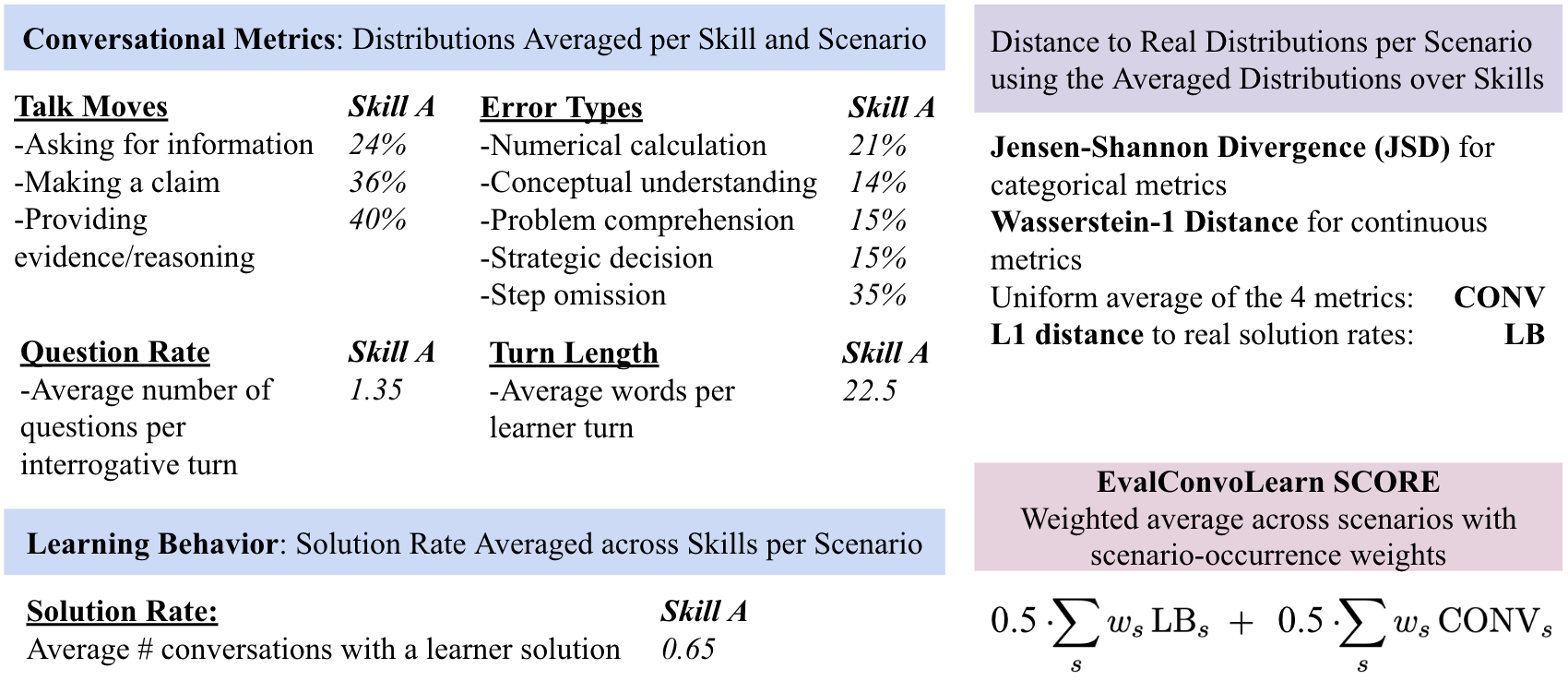}
    \caption{EvalConvoLearn metrics aggregation process}
    \label{fig:metrics_aggregation}
\end{figure}

Following \citet{Scarlatos2025} applying Knowledge Tracing models \citep{corbett1994knowledge} to tutoring dialogues at the turn level, we treat each learner turn as a skill practice attempt. We cap conversations at $7$ turns to allow time for learning while preventing unrealistic lengths, following a heuristic of seven attempts to mastery \citep{koedinger2023}. We strategically select a \emph{skill pool} with sufficient aligned dialogues in the dataset that support all mastery scenarios. The learning-behavior score $\text{LB}$ is the L1 distance between simulated and real outcome distributions for each scenario and skill, macro-averaged across the skill pool (Figure \ref{fig:metrics_aggregation}).

We evaluate conversational realism with four metrics: The presence of three talk moves \citep{suresh2022talkmovesdatasetk12mathematics} and five error types \citep{qi_simulating_2026} (LLM-labeled), number of questions in interrogative learner turns \citep{perczel2025teachlmposttrainingllmseducation}, and turn length \citep{perczel2025teachlmposttrainingllmseducation}. We manually evaluated LLM labels on 50 conversations and got a per-label Cohen's $\kappa$ between .73 and 1.0 for talk moves (exact match .96) and from .79 to 1.0 for error types (exact match .88).
We score each metric by computing the distance between its distribution in the simulated and real data within each scenario, using Jensen–Shannon divergence (JSD) or Wasserstein-1 distance (Figure \ref{fig:metrics_aggregation}). The resulting conversational score $\text{CONV}_s$ averages the four distance metrics with equal weights. Users may add other conversational metrics and weigh them differently for their own context and evaluation goals.

The final $\text{EvalConvoLearn}$ score (ECL) averages $\text{CONV}_s$ and $\text{LB}_s$ across scenarios $s$, accounting for scenario distributions from their occurrence in real dialogues (Figure \ref{fig:metrics_aggregation}). Both scores may be weighted according to their relative importance in one's learning context.

\section{Evaluation on the Eedi Dataset}
\label{sec:eedi}

We apply EvalConvoLearn to the largest publicly available dataset of student–TA tutoring dialogues, from the Eedi learning platform~\citep{zent-etal-2025-piivot}. Since students proactively seek help on Eedi, we assume they have \emph{not} yet mastered the target item skill but \emph{have} mastered its prerequisites given Eedi's progressive curriculum. This assumption could be relaxed with random item assignments to learners, e.g. in simulated datasets. We retain conversations with $\geq$50\% student utterances, conducted by tutors with $\geq$5 conversations (to enable few-shot prompting), yielding 66 conversations. We tag each practice item with a single skill from a subset of the B.E.S.T.\footnote{\url{https://www.fldoe.org/academics/standards/subject-areas/math-science/mathematics/}} curriculum using GPT-4.1-mini for its cost/accuracy trade-off, and match simulations by keeping the $7$ first conversation turns only.

We simulate conversations between a tutor using 3-shot prompting and two LLM-based learners using GPT-4.1-mini with custom knowledge implementations: The \textbf{Conversation Summaries} learner stores summaries of past conversations and uses them as context when generating a new response. The \textbf{Binary Skills} learner uses LLM-as-a-judge to tag items with skills and stores binary mastery outcomes after conversations. Table~\ref{tab:results} reports Conversational (CONV), Learner Behavior (LB), and ECL scores for each configuration, aggregated across 3 skills with 8 conversations per skill. We used Claude's Sonnet-4.6 for tutor responses and metric evaluations.

\begin{table}[htbp]
\floatconts
  {tab:results}
  {\caption{EvalConvoLearn results across 3 runs (mean $\pm$ standard error). Lower is closer to real distributions. The first row reports the CONV breakdown distances; the second row reports the aggregated scores (CONV, LB, ECL).}}
  {\begin{tabular}{l cccc}
  \toprule
  \textbf{Setting} & \textbf{Talk Moves}   & \textbf{Errors} & \textbf{Questions}     & \textbf{Turn Length} \\
                   & \textbf{CONV} & \textbf{LB}  & \textbf{ECL} &              \\
  \midrule
\multirow{2}{*}{Binary Skills} & 0.207 $\pm$ 0.000 & \textbf{0.385 $\pm$ 0.042} & 0.415 $\pm$ 0.021 & 0.724 $\pm$ 0.022 \\
  & \textbf{0.433 $\pm$ 0.011} & \textbf{0.556 $\pm$ 0.014} & \textbf{0.494 $\pm$ 0.009} &  \\
\addlinespace
\multirow{2}{*}{Conv. History} & \textbf{0.206 $\pm$ 0.006} & 0.708 $\pm$ 0.025 & \textbf{0.226 $\pm$ 0.006} & \textbf{0.658 $\pm$ 0.011} \\
  & 0.450 $\pm$ 0.004 & 0.556 $\pm$ 0.014 & 0.503 $\pm$ 0.008 &  \\
  \bottomrule
  \end{tabular}}
\end{table}

\section{Discussion and Future Work}
\label{sec:discussion}
The learners have similar $LB$ scores as they both solve around 90\% of items, which is 56 points more than real students, but differ slightly in conversational metrics while keeping high within-learner consistency. These results provide a signal to iterate on each learner's design by targeting their conversational or learning abilities, and suggest a large room for improvement for the models to produce realistic learner behavior. Future improvements should also tune the tutor model (See our ablation study of tutor models \ref{apd:skills}), moving towards a dual learner-tutor optimization task.

EvalConvoLearn is a work in progress. Critically, we are looking to expand the suite of scenarios and further validate the metrics with human ratings. When released, users can test the framework with new datasets, advanced learner and tutor simulations (e.g. fine-tuned or RL-trained models~\citep{Scarlatos2025}), and iterate on the learner quality signal to develop learner simulation benchmarks for various educational contexts.

\acks{The author thanks the Learning Engineering Virtual Institute, a program of Renaissance Philanthropy, for supporting the development of this work, as well as AJ Strauman-Scott for her contribution to the project.}

\bibliography{evalconvolearn}

\appendix
\section{Tutor Context Ablation study}
\label{apd:skills}
Because learner behavior depends on tutor messages, ECL scores can only be compared across simulations with similar tutor behavior distributions. We attempt to reduce the variance on the tutor's behavior and make it more realistic using few-shot prompting with individual tutor responses extracted from the dataset.
To measure the impact of different tutors on the EvalConvoLearn scores, we conducted an ablation study by removing sample tutor responses from tutor prompts, with results in Table 2.

\begin{table}[htbp]
\floatconts
  {tab:ablation}
  {\caption{Simulated learner scores with and without tutor prompting contexts, for Binary Skills (BS) and Conversation Summaries (CS) learners}}
  {\begin{tabular}{lllccc}
  \toprule
  \textbf{Learner} & \textbf{Eval} & \textbf{Tutor} & \textbf{CONV} & \textbf{LB} & \textbf{ECL} \\
  \midrule

BS, GPT-4.1-mini & Sonnet 4.6 & 3-shot & 0.433 $\pm$ 0.011 & \textbf{0.556 $\pm$ 0.014} & \textbf{0.494 $\pm$ 0.009} \\
 BS, GPT-4.1-mini & Sonnet 4.6 & 0-shot & \textbf{0.406 $\pm$ 0.004} & 0.611 $\pm$ 0.014 & 0.509 $\pm$ 0.007 \\
\midrule
CS, GPT-4.1-mini & Sonnet 4.6 & 3-shot & 0.450 $\pm$ 0.004 & \textbf{0.556 $\pm$ 0.014} & 0.503 $\pm$ 0.008 \\
CS, GPT-4.1-mini & Sonnet 4.6 & 0-shot & 0.453 $\pm$ 0.035 & 0.653 $\pm$ 0.014 & 0.553 $\pm$ 0.025 \\
  \bottomrule
  \end{tabular}}
\end{table}

We observe that adding few-shot prompting improves final ECL scores for both learners, mainly through consistently better learner behavior. However, conversational scores see no change or get worse (in the case of the Binary Skills learner) when adding few-shot prompting. Adding human tutor examples seems to reduce the propensity of learners to solve problems in $7$ conversation turns, which may be explained by shorter, more conversational, or less information-rich tutor responses. Simple analysis (N=450) corroborates this hypothesis by showing longer average tutor responses in the 0-shot setting (51 words) compared to the 3-shot setting (46 words), which is expected as few-shot prompting tends to correct the LLM propensity to be too verbose \cite{nayab2024concise}.

Additionally, the observed difference between Binary Skills learners' $CONV$ scores comes specifically from more realistic error types in the 0-shot setting than the 3-shot setting (JSD of .29 versus .39, not reported here). Further work is needed to analyze the differences of tutor response features across few-shot settings, and refine the "error types" metric in particular by understanding how different tutor-learner conversational interactions can shift the score's distribution.

\section{LLM Prompts}
\label{apd:prompts}
All code and prompts are available on the public GitHub code.

\end{document}